\documentclass[conference]{IEEEtran}
\IEEEoverridecommandlockouts
\usepackage{amsmath,amssymb,amsfonts}
\usepackage{bbm}
\usepackage{graphicx}
\graphicspath{{./figs/}}
\usepackage{bm}
\usepackage{cite}
\usepackage{comment}
\usepackage{xcolor}
\usepackage{url}
\usepackage{mathrsfs}
\usepackage{caption}
\usepackage{subcaption}
\usepackage{float}
\usepackage{xcolor}
\usepackage{booktabs}
\usepackage{slashbox}
\usepackage[hidelinks]{hyperref}
\usepackage[utf8]{inputenc}
\usepackage{mathrsfs}
\usepackage{tikz}
\usepackage{pgfplots}
\usepgfplotslibrary{patchplots} % 
\pgfplotsset{compat=1.18}
\usepackage{adjustbox}
\usepackage[linesnumbered,ruled,vlined]{algorithm2e}
\usepackage{booktabs} % For professional-quality tables

\begin{document}

\title{Tracking Time-Varying Multipath Channels for Active Sonar Applications 
\thanks{This work was partially supported by the Wallenberg AI, Autonomous Systems and
Software Program (WASP) funded by the Knut and Alice Wallenberg Foundation.}
}
\author{%
\IEEEauthorblockN{Ashwani Koul$^{\dagger}$,
Gustaf Hendeby$^{\dagger}$,
Isaac Skog$^{\ddagger}$}
\IEEEauthorblockA{%
$^{\dagger}$Division of Automatic Control, Link\"oping University, Link\"oping, Sweden\\
\{ashwani.koul, gustaf.hendeby\}@liu.se\\[0.25em]
$^{\ddagger}$Division of Communication Systems, KTH Royal Institute of Technology, Stockholm, Sweden, and\\
FOI-Swedish Defence Research Agency, Stockholm, Sweden\\
skog@kth.se
}}

\maketitle
\begin{abstract}
Reliable detection and tracking in active sonar require accurate and efficient learning of the acoustic multipath background environment. Conventionally, background learning is performed after transforming measurements into the range–Doppler domain, a step that is computationally expensive and can obscure phase-coherent structure useful for monitoring and tracking. This paper proposes a framework for learning and tracking the multipath background directly in the raw measurement domain. Starting from a wideband Doppler linearization of the impulse response of a time-varying multipath channel, a state-space model with a heteroscedastic measurement equation is derived. This model enables channel tracking using an extended Kalman filter (EKF), and unknown model parameters are learned from the marginalized likelihood. The statistical adequacy of the proposed models is assessed via a $p$-value significance test. Finally, this paper integrates the learned channel model into a sequential likelihood-ratio test for target detection. BELLHOP-based simulations show that the proposed model better captures channel dynamics induced by sea-surface fluctuations and transmitter and receiver drift, yielding more reliable detection in time-varying shallow-water environments.
\end{abstract}


\section{Introduction}
Accurate modeling of the multipath background is essential for reliable active-sonar detection and tracking. In shallow waters, repeated surface and bottom interactions produce strong reverberation and coherent multipath propagation whose statistics vary over time due to surface dynamics and platform drift. This nonstationarity makes estimating multipath background difficult, increasing false-alarm rates, and masking weak target returns. This, in turn, makes reliable detection and tracking difficult in low signal-to-noise ratio (SNR) regimes~\cite{Willis2005, Abraham2019}. 

Classical sonar pipelines map the received time series to range-Doppler via matched filtering (often with Doppler filter banks) and apply constant false alarm rate (CFAR)-type thresholding~\cite{Kalyan, Delyon2020, Bald2006}. Their performance relies on the interference statistics being approximately stationary over the CFAR training window, an assumption that can break down under time-varying, reverberation-dominated multipath, leading to threshold mismatch and degraded detection.

\begin{figure}[t]
    \centering
    \includegraphics[width=0.5\textwidth]{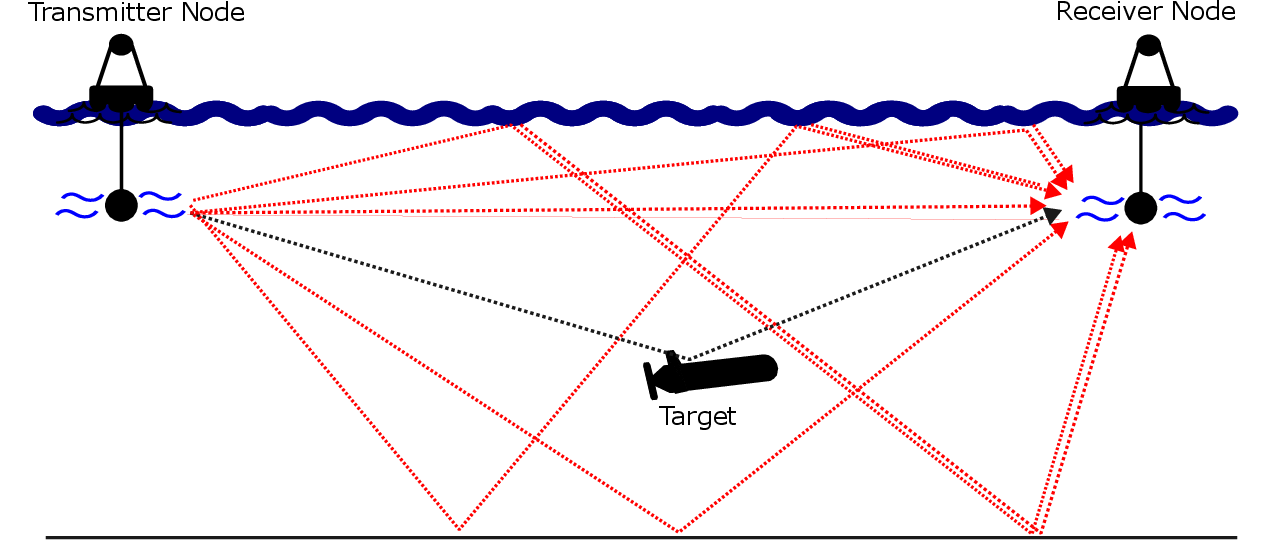}
    \caption{Illustration of the considered bistatic sonar scenario with a time-varying multipath channel.}
    \label{figsetup}
\end{figure}

For bistatic sonar, as illustrated in Fig.~\ref{figsetup}, the structured multipath background can evolve due to sensor-node drift and surface motion~\cite{Abraham2019,miliucom}. Under such conditions, conventional range-Doppler processing requires frequent matched filtering, which causes a computational burden. In addition, Doppler mismatch and windowing can spread coherent multipath energy across range-Doppler bins, making background evolution harder to model. These effects motivate operating directly on raw measurements and explicitly tracking the time-varying multipath background.

Instead of operating in the range-Doppler domain, several approaches work directly with raw sensor measurements, enabling likelihood-based formulations such as track-before-detect, and bearing-only tracking using the maximum a posteriori (MAP) method. Operating directly with raw measurements has shown improved performance in low-SNR regimes~\cite{Bosser2025,Kullberg2025}. Moreover, raw-domain processing retains the phase-coherent structure of the received field. 

In~\cite{Yang2024}, weak targets are detected by tracking the time-varying channel impulse response (CIR) using a block-updated sparse estimator. Even though such models achieve good performance, in high-reverberant environments with no structure and sparsity, model mismatch will limit performance. It can favor faster, non-structured adaptive updates, resulting in fewer target detections.

In this paper, a different approach for target detection is proposed. Instead of estimating the CIR and then computing the residuals for target detection, the time-varying multipath channel is tracked using an extended Kalman Filter (EKF). Using a wideband Doppler linearization approximation, the multipath background is modelled directly using the raw measurements. With this, a detector is designed for weak target detection based on the changes in the marginal likelihood of the measurements~\cite{Skog2024, Abraham2019}. Tracking the multipath channel is done by modeling the time-varying components using a heteroscedastic measurement model for the sensor measurements. Different models are proposed, and the significance of each of the models is ascertained by a $p$-value significance test. Hence, the resulting framework enables simultaneous tracking of the time-varying multipath background and detection of a target. 

To that end, the main contributions of this paper are:
\begin{itemize}
    \item A raw-measurement-domain background model for active sonar in which Doppler effects are captured through a physically interpretable, state-dependent covariance structure.
    \item  A EKF-based framework for tracking the multipath channel, and learning hyperparameters in the proposed channel model. 
    \item A simulation-based evaluation demonstrating the statistical adequacy of the proposed background model via $p$-value significance testing, and its impact on detection performance in time-varying environments.
\end{itemize}

\noindent\textbf{Reproducible research:} The code and datasets used to generate the results in this paper are available at \\ \texttt{\url{https://github.com/ASHKoul/Bcg_SLRT}}.

\section{Signal Model}\label{sec:bg_model}
Consider the bistatic setup in Fig.~\ref{figsetup}. The received data consist of a time-varying multipath background and a possible target return. To enable likelihood-based target detection and tracking, the modeling of the multipath background, its statistical properties, and its time evolution has to be performed.
\subsection{General Continuous Time Channel Model}
 Let $s(t)$ denote the transmitted waveform. A standard continuous-time multipath model for the received signal, in a noise-free environment, is
\begin{equation}\label{eq:tv_multipath_basic}
x(t)=\sum_{i=1}^{N_p} a_i(t)\,s\big(t-\tau_i(t)\big), \qquad t\in[0,t_f],
\end{equation}
where $\tau_i(t)$ and $a_i(t)$ are the time-varying delay and amplitude of the $i^{th}$ arrival, $N_p$ is the number of significant arrivals. In shallow water, these arrivals include the direct path and strong surface and bottom interactions, yielding a reverberation-dominated multipath background in addition to the target returns.

Assuming the channel varies slowly over the time duration $t_f$, the multipath amplitude and delay can be approximated as
\begin{equation}\label{eq:linear_tau}
a_i(t)\approx a_i,\qquad \tau_i(t)\approx \tau_{io}+\eta_i t,\qquad t\in[0,t_f],
\end{equation}
where $\tau_{io}$ is the initial delay and $\eta_i$ is the delay-rate.\\
Then
\begin{equation}
 t-\tau_i(t) \approx (1-\eta_i)t-\tau_{io}=\beta_i\left(t-\bar\tau_i\right),   
\end{equation}
with
\begin{equation}\label{eq:beta_taubar}
\beta_i\triangleq 1-\eta_i \quad\textrm{and} \quad \bar\tau_i\triangleq \frac{\tau_{io}}{\beta_i}.
\end{equation}
Substituting~\eqref{eq:beta_taubar}  into~\eqref{eq:tv_multipath_basic}, $x(t)$ can be approximated as
\begin{equation}\label{eq:tv_multipath_scaled}
x(t) \approx \sum_{i=1}^{N_p} a_i\, s\big(\beta_i(t-\bar\tau_i)\big).
\end{equation}

To separate the potential target components from the multipath background,~\eqref{eq:tv_multipath_scaled} can be split as
\begin{subequations}
\begin{equation}\label{eq:y_split}
x(t)=x_b(t)+x_o(t),
\end{equation}
where 
\begin{equation}\label{eq:bcpart}
    x_b(t)\triangleq\sum_{i=1}^{N_b} a_i\, s\big(\beta_i(t-\bar\tau_i)\big)
\end{equation}
denotes the multipath background, and 
\begin{equation}\label{eq:targetpart}
    x_o(t)\triangleq\sum_{i={N_b+1}}^{N_p} a_i s(\beta_i(t-\bar\tau_i))
\end{equation} 
is the multipath structure formed due to the target-reflected signal. Here, for notational convenience, the arrivals are ordered such that $N_b$ and $N_o$ arrivals are attributed to background and target-induced components, respectively, with $N_p=N_b+N_o$.
\end{subequations}
Since the target amplitudes are much smaller relative to the multipath background, the target multipath is approximated by its dominant target arrival as
\begin{equation}\label{eq:target_approx}
    x_o(t;\bar{\tau}_o,\beta_o) \approx a_o s(\beta_o(t-\bar{\tau}_o)).
\end{equation}

\subsection{Wideband Linearized Model}
The multipath background can be further approximated using a first-order wideband Doppler linearization (see Appendix~\ref{app:linearization}),
\begin{subequations}
\begin{align}\label{eq:bg_lin}
x_b(t)\approx &\sum_{i=1}^{N_b} \Big( a_i\, s(t-\bar\tau_i)\;+\; a_i\,r_i\,(t-\bar\tau_i)\dot s(t-\bar\tau_i) \Big),\\
=& \sum_{i=1}^{N_b} \Big( a_i\, s(t-\bar\tau_i)\;+\; a_i\,r_i\,u(t-\bar\tau_i) \Big),
\end{align}
\end{subequations}
where $r_i\triangleq \ln\beta_i$ and $ u(t)\triangleq t\,\dot{s}(t)$. This approximation is accurate when $\beta \approx 1$.
Equation~\eqref{eq:bg_lin} is bi-linear in the unknown parameters $a_i$ and $r_i$ and can be interpreted as a linear convolutional model driven by the two known waveforms $s(t)$ and $u(t)$. This simplifies estimation and tracking of the time-varying multipath background.

\subsection{Discrete-Time Linearized Channel Model}
Defining the sampled transmitted waveform and the corresponding time derivative component with the sampling period  $\Delta_t$ as
\begin{equation}\label{eq: sampled signal}
s[n]\triangleq s(n\Delta_t) \quad \textrm{and}\quad u[n]\triangleq n\Delta_t\,\dot s(n\Delta_t).
\end{equation}
With this, the zero-padded Toeplitz matrices $S$ and $U$ of size $N\times N_l$ can be defined as
\begin{equation}
[S]_{n,l}\triangleq s[n-l] \quad \textrm{and}\quad [U]_{n,l}\triangleq u[n-l],
\end{equation}
where $n=0,\cdots, N-1$, $l=0,\cdots, N_l-1$, $N\triangleq \lceil t_f/\Delta_t\rceil$, and $ N_l$ is the length of the sampled delay grid. Given these quantities and defining 
$\vec{x}_b \triangleq \begin{bmatrix} x_b(0) &\cdots& x_b((N-1)\Delta_t) \end{bmatrix}^\top$,
a discrete time approximation in vector form of the time continuous model~\eqref{eq:bg_lin} is given by
\begin{equation}\label{eq:bg_vec}
\vec{x}_b \approx Sa + U\,\Lambda_a\, r,
\end{equation}
where $a=\begin{bmatrix} a_1 & \cdots & a_{N_l} \end{bmatrix}^\top$ and $r=\begin{bmatrix} r_1 & \cdots & r_{N_l}\end{bmatrix}^\top$ are $N_l$ ($ N_l \gg N_p$) length vectors collecting the multipath amplitudes and log time-scaling parameters, respectively. Further, $\Lambda_a\triangleq\mathrm{diag}(a)$.

\subsection{Stochastic Multipath Background Model}
The environmental fluctuations captured by the time scaling parameter $r$ can be decomposed into
\begin{equation}
r \triangleq c+d,
\end{equation}
where $c$ and $d$ model the Doppler effect unique to each signal path and the common Doppler effects, respectively.
Assume that $c$ and $d$ are zero-mean random variables with the distribution
\begin{equation}
    c\sim\mathcal{N}(0,\sigma_c^2 I), \quad \text{and} \quad d\sim\mathcal{N}(0,\sigma_d^2 \mathbbm{1}\mathbbm{1}^\top),
\end{equation}
where $\mathbbm{1}=[1~\cdots~1]^\top$. Then the measurements, conditioned on the amplitude vector $a$, can be modeled as
\begin{equation}
y_b = Sa +U\Lambda_a r + e,\qquad e\sim\mathcal{N}(0,\sigma_e^2 I),
\end{equation}
where $\sigma_e^2$ denotes the power of the ambient noise, $y_b$ is the collection of discrete-time measurements when the target is absent, and 
\begin{equation}\label{eq:yb_cond_a}
y_b\,|\,a \sim \mathcal{N}(Sa,\; R(a)),\qquad
R(a)\triangleq \sigma_e^2 I + \Sigma_b(a),
\end{equation}
with
\begin{equation}\label{eq:Sigma_b_a}
\Sigma_b(a)= \Sigma_c (a) +\Sigma_d(a).
\end{equation}
Here, 
\begin{equation}\label{eq:Sigma_c_d}
\begin{aligned}
\Sigma_c(a) \triangleq & \,\,\sigma_c^2 \,U \Lambda_a\Lambda_a^\top U^\top,\\
\text{and}\quad \Sigma_d(a) \triangleq & \,\,\sigma_d^2 \,Uaa^\top U^\top.
\end{aligned}
\end{equation}
are the heteroscedastic covariance matrices, and the measurement model becomes heteroscedastic because $R(a)$ depends on the unknown background amplitudes $a$.

\subsection{State-Space Model}
Due to surface dynamics and sensor nodes' drift, the multipath amplitudes vary from ping to ping. Moreover, in shallow water, the multipath background is typically smooth or diffused in delay, and estimating a full $N_l$-tap vector $a$ becomes computationally expensive when $N_l\gg N_p$. Therefore, a low-dimensional parameterization is imposed by defining
\begin{equation}\label{eq:a_basis_theta}
a \triangleq B\theta,
\end{equation}
where $\theta$ is a length-$M$ vector of unknown basis weights and $B$ is a fixed $N_l\times M$ basis dictionary. The dictionary $B$ is constructed using Gaussian basis functions,
\begin{equation}
[B]_{l,m}\triangleq \exp\!\left(-\frac{(l\Delta_t-\mu_m)^2}{2\sigma_m^2}\right),
\end{equation}
where $\mu_m$ and $\sigma_m$ is the center and the length scale of the $m^{th}$ basis, respectively. The centers $\mu_m$ are placed uniformly over the delay grid, whereas the length scale is set $\sigma_m \approx 0.42/BW$, with $BW$ as the bandwidth of the transmitted signal. 

Substituting~\eqref{eq:a_basis_theta} into~\eqref{eq:yb_cond_a} gives the measurement model
\begin{equation}\label{eq:reparam_stat_model_clean}
y_b \,|\, \theta \sim \mathcal{N}\!\big(H\theta,\;R(\theta)\big),
\qquad H\triangleq SB,
\end{equation}
where $R(\theta)$ can be computed by substituting $a=B\theta$ in~\eqref{eq:yb_cond_a}.

With this, let $\theta_k$ and $y_{b,k}$ denote the basis coefficient and the measurement vector at ping $k$, respectively. The evolution of the basis coefficients across pings is modeled as a random-walk process. The resulting state-space model under the background-only hypothesis is given as
\begin{subequations}\label{eq:ssm_clean}
\begin{align}
\theta_{k+1} &= \theta_k + w_k,\qquad w_k\sim \mathcal{N}(0,\sigma_q^2 I),\\
y_{b,k} &= H\theta_k + v_k,\qquad v_k\sim \mathcal{N}(0,R(\theta_k)), \label{eq:mult_meas}
\end{align}
\end{subequations}
where $w_k$ and $v_k$ are mutually independent, and $v_k$ captures both ambient noise and state-dependent background fluctuations with $R(\theta_k)=\sigma_e^2 I +\Sigma_b(\theta_k)$. 

\section{Parameter Estimation}
The state-space model in~\eqref{eq:ssm_clean} enables sequential estimation of the latent multipath background state $\theta_k$, which describes the multipath channel.
\subsection{Amplitude Tracking}
The time-varying multipath channel amplitudes $a$ are estimated by tracking the low-dimensional coefficient vector $\theta_k$. The tracking of $\theta_k$ is performed using an EKF, as outlined in Alg.~\ref{alg:mle_plugin_ekf}. The measurement covariance $R(\theta_k)$ is state-dependent due to the heteroscedastic noise model. To obtain a tractable update, a locally constant approximation of the covariance is used by evaluating the covariance at the predicted latent state $\hat{\theta}_{k|k-1}$, i.e.,
\begin{equation}
R_k \triangleq R(\hat{\theta}_{k|k-1}).
\end{equation}
\subsection{Hyperparameter Learning}
Not all the parameters, such as $\{\sigma_q, \sigma_c, \sigma_d\}$, in the model~\eqref{eq:ssm_clean} may be known before hand. These parameters, here denoted by $\Theta$, can be learned from the data by maximizing the log marginal likelihood~\cite{Durbin2012}. That is
\begin{equation}
\hat{\Theta}=\arg\max_{\Theta} 
L(Y_{b,N_{k}};\Theta),
\end{equation}
where $Y_{b,N_{k}}=\{y_{b,k}\}_{k=1}^{N_{k}}$. Here, the approximation of $L(Y_{b,N_{k}};\Theta)$ can be calculated via the EKF; see Alg.~\ref{alg:mle_plugin_ekf}.

\begin{algorithm}[t]
\caption{EKF for estimating the amplitude $\theta$ and calculating the marginal likelihood.}
\label{alg:mle_plugin_ekf}
\DontPrintSemicolon
\SetKwInOut{Input}{Input}
\SetKwInOut{Output}{Output}
\SetKwInOut{Initialize}{Initialize}

\Input{Measurements $\{y_k\}_{k=1}^{N_{k}}$, measurement matrix $H$, hyperparameter set $\Theta$, model $\mathcal{M}$.}
\Initialize{$(\hat{\theta}_{0|0},P_{0|0})$}
\BlankLine
\tcp{For model $\mathcal{M}$}
    \For{$k=1,\dots,N_{k}$}{
        \tcp{Time update}
        $\hat{\theta}_{k|k-1} \leftarrow \hat{\theta}_{k-1|k-1}$\;
        $P_{k|k-1} \leftarrow P_{k-1|k-1} + \sigma_q^2 I$\;
        \tcp{State-dependent covariance}
        $R_k \leftarrow R_\mathcal{M}\!\big(\hat{\theta}_{k|k-1}\big)$\;

        \tcp{Measurement update}
        $\nu_k \leftarrow y_k - H\hat{\theta}_{k|k-1}$\;
        $\Sigma_k \leftarrow H P_{k|k-1} H^\top + R_k$\;
        $K_k \leftarrow P_{k|k-1} H^\top \Sigma_k^{-1}$\;
        $\hat{\theta}_{k|k} \leftarrow \hat{\theta}_{k|k-1} + K_k \nu_k$\;
        $P_{k|k} \leftarrow (I-K_k H)P_{k|k-1}$\;

        \tcp{Log-likelihood increment}
        $\ell_k(y_k;\Theta) \triangleq \log p\!\left(y_k \mid y_{1:k-1};\Theta\right)$\;
        $\qquad \qquad
        \approx -\frac{1}{2}\!\left(\nu_k^\top \Sigma_k^{-1}\nu_k + \log|\Sigma_k|\right)$\;
    }
    $L(Y_{N_{k}};\Theta) =\sum_{k=1}^{N_{k}} \ell_k(y_k;\Theta)$\;
\Output{posterior estimates $\{\hat{\theta}_{k|k},P_{k|k}\}_{k=1}^{N_{k}}$, and log marginal likelihood $L(Y_{N_{k}};\Theta)$.}
\end{algorithm}

\section{Evaluation Methodology}\label{sec:proposed_bg_models}
A central question is whether explicitly modeling time-varying background fluctuations yields a statistically significant improvement over an ambient noise-only baseline, i.e., $\sigma_c=\sigma_d=0$, and whether this translates into more reliable target detection. To that end, model significance and target detection capability are evaluated using the following four covariance structures
\begin{subequations}\label{eq:models_clean}
\begin{align}
\mathcal{M}_0:\;& R_{\mathcal{M}_0}(\theta_k)=\sigma_e^2 I,\\
\mathcal{M}_c:\;& R_{\mathcal{M}_c}(\theta_k)=\sigma_e^2 I + \Sigma_c(\theta_k),\\
\mathcal{M}_d:\;& R_{\mathcal{M}_d}(\theta_k)=\sigma_e^2 I + \Sigma_d(\theta_k),\\
\mathcal{M}_{cd}:\;& R_{\mathcal{M}_{cd}}(\theta_k)=\sigma_e^2 I + \Sigma_c(\theta_k)+\Sigma_d(\theta_k),
\end{align}
\end{subequations}

\subsection{$P$-value Significance Test}\label{subsec:pvalue}
Let $\mathcal{M}\in\{\mathcal{M}_0,\mathcal{M}_c,\mathcal{M}_d,\mathcal{M}_{cd}\}$ index the candidate covariance model for $R(\theta_k)$ in~\eqref{eq:mult_meas}, and
\begin{equation}
y_{b,k}=H\theta_k + v_{\mathcal{M},k},\qquad
v_{\mathcal{M},k}\sim \mathcal{N}(0,R_{\mathcal{M}}(\theta_k)).
\end{equation}

Then the log likelihood-ratio statistic for each extended model $\mathcal{M}_j$, $j\in \{c, d, cd\}$ relative to $\mathcal{M}_0$ is
\begin{subequations}\label{eq:LLR_clean}
\begin{align}
T_j(Y_{b,N_{k}})
&\triangleq \log \frac{\underset{\Theta_j}{\max}\, p(Y_{b,N_{k}};\Theta_j,\mathcal{M}_j)}
{\underset{\Theta_0}{\max}\, p(Y_{b,N_{k}};\Theta_0,\mathcal{M}_0)}\\
&= L(Y_{b,N_{k}};\hat\Theta_j,\mathcal{M}_j)-L(Y_{b,N_{k}};\hat\Theta_0,\mathcal{M}_0).
\end{align}
\end{subequations}
The hyperparameter set for each model is
\begin{align*}
\Theta_{cd}&=\begin{bmatrix}\sigma_q^2 & \sigma_c^2 & \sigma_d^2\end{bmatrix}^\top,\\
\Theta_{c}&=\begin{bmatrix}\sigma_q^2 & \sigma_c^2\end{bmatrix}^\top,\\
\Theta_{d}&=\begin{bmatrix}\sigma_q^2 & \sigma_d^2\end{bmatrix}^\top,\\
\Theta_0&=\sigma_q^2,
\end{align*}
where $\sigma_q^2$ is present under all models and $\sigma_c^2$ or $\sigma_d^2$ augment the covariance structure relative to $\mathcal{M}_0$. The noise variance $\sigma_e^2$ is assumed known from data segments with no transmitted pulse.
\begin{figure*}[t]
    \centering
    \begin{subfigure}[t]{0.33\textwidth}
        \centering
            \includegraphics[width=1\textwidth,height=0.7\textwidth]{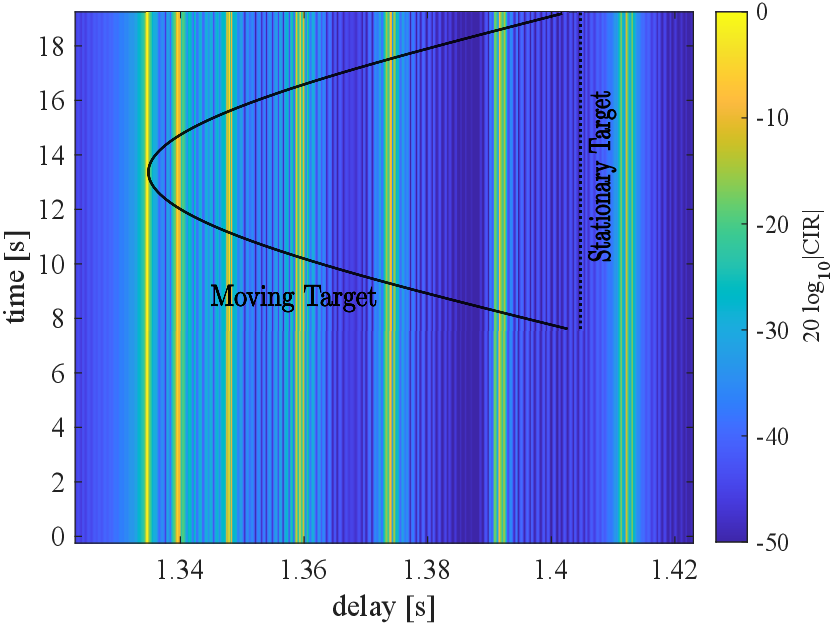}
        \caption{Scenario 1.}
        \label{fig:cir_static}
    \end{subfigure}\hfill
    \begin{subfigure}[t]{0.33\textwidth}
        \centering
            \includegraphics[width=1\textwidth,height=0.7\textwidth]{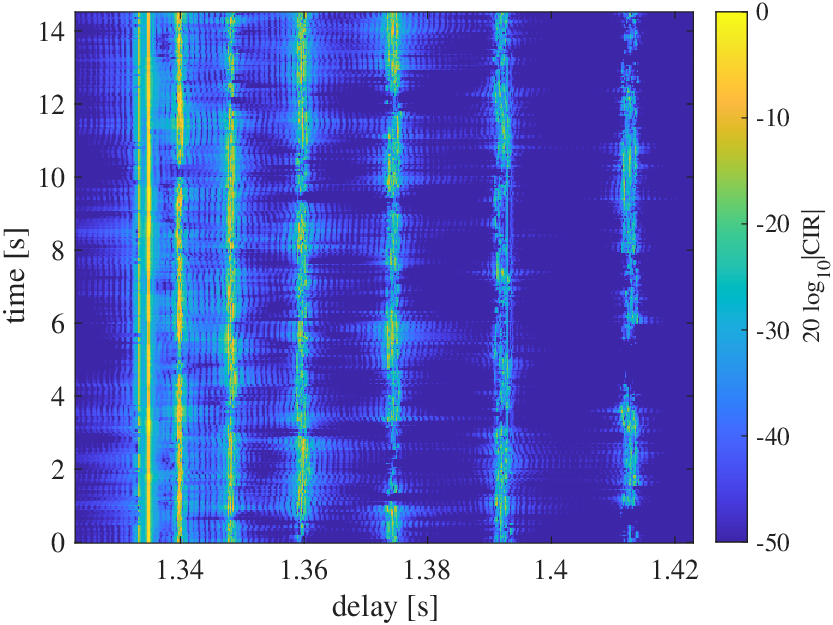}
        \caption{Scenario 2.}
        \label{fig:cir_reverb_only}
    \end{subfigure}
    \begin{subfigure}[t]{0.33\textwidth}
        \centering
            \includegraphics[width=1\textwidth,height=0.7\textwidth]{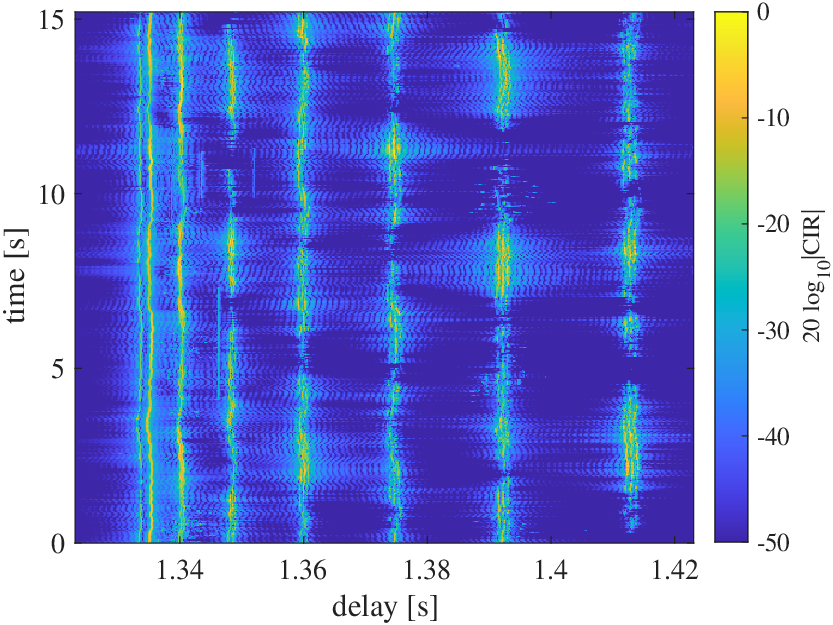}
        \caption{Scenario 3.}
        \label{fig:cir_drift_reverb}
    \end{subfigure}
    \caption{Simulated CIRs for three scenarios. The trajectories for the stationary and moving target are also shown in (a).}
    \label{fig:cir_bellhop_scenarios}
\end{figure*}

Using Wilks' theorem, the distribution of the test statistics under the null hypothesis is approximated as
\begin{equation}
2T_j(Y_{b,N_{k}})\mid _{\mathcal{M}_0}\ \overset{a}{\sim}\ \chi^2_{\zeta_j},\quad 
\zeta_j=\dim(\Theta_j)-\dim(\Theta_0).
\end{equation}
Hence, the  $p$-value is computed as
\begin{equation}\label{eq: ptest}
p_j \;=\; \Pr\!\left(\chi^2_{\zeta_j}\ge 2T_j\right)
\end{equation}
 with $\mathcal{M}_0$ being preferred to $\mathcal{M}_j$ unless $p_j\le \alpha$.\\
\subsection{Target Detection Performance}\label{subsec:slrt}
Let $\vec{x}_{o,k}$ represent a collection of discrete-time samples for the target reflection during the $k^{th}$ ping, and is defined as
\begin{equation}
    \vec{x}_{o,k}=\begin{bmatrix}x_{o,k}[0] & \cdots & x_{o,k}[N-1] \end{bmatrix}^\top,
\end{equation}
where \begin{equation}
x_{o,k}[n] =
\begin{cases}
0, & k<\bar{k}\\
a_{o,k}\,s(\beta_{o,k}(n\Delta_t -\bar{\tau}_{o,k})), & k\ge \bar{k}
\end{cases}
\end{equation}
and $\{a_{o,k},\bar{\tau}_{o,k},\beta_{o,k}\}$ are the target parameters during the $k^{th}$ ping.
Furthermore, $\bar{k}$ represents the unknown onset ping index where the target is assumed to appear. 
With this, the discrete-time measurement vector is given as
\begin{equation}
y_k = H\theta_k +\vec{x}_{o,k}+v_{k}.
\end{equation}
Our goal is to quantify the benefit of explicitly modeling and tracking the multipath background before performing target detection. Since the target onset time is unknown, a sequential likelihood ratio test (SLRT) can be employed under different models ($\mathcal{M}_0,\mathcal{M}_c,\mathcal{M}_d,\mathcal{M}_{cd}$) and evaluate the resulting detection performance.

To that end, let $\mathcal{H}_0$ and $\mathcal{H}_1$ denote the background-only and target-present hypotheses for a given model $\mathcal{M}$:
\begin{subequations}
\begin{align}
\mathcal{H}_0:\;\; y_k &= H\theta_k + v_{\mathcal{M},k}, \label{eq:detH0}\\
\mathcal{H}_1:\;\; y_k &= H\theta_k + \vec{x}_{o,k} + v_{\mathcal{M},k}, \label{eq:detH1}
\end{align}
\end{subequations}
Here, the focus is on quantifying the detection benefit for different background models, and therefore the target waveform parameters $(\bar{\tau}_{o,k},\beta_{o,k})$ and amplitude $a_{o,k}$ are assumed known; only the onset ping index is unknown.

Now, the sequential log-likelihood ratio is computed as
\begin{subequations}
\begin{equation}
G_{k+1}=G_k+\gamma_k,\quad \forall k=k_o,k_o+1,\ldots
\end{equation}
with
\begin{equation}
\gamma_k \triangleq 
\log \frac{p(y_k \mid y_{k_o:k-1};\hat{\Theta},\mathcal{H}_1)}{p(y_k \mid y_{k_o:k-1};\hat{\Theta},\mathcal{H}_0)} ,
\end{equation}
where $G_{k_o}=0$ and $k_o$ denotes the ping at which the test is (re)started, and the marginal likelihoods are calculated via the EKF. Here, the notation $\mathcal{M}$ is suppressed as the model remains the same in both hypotheses. The stopping rule is given as
\begin{equation}
\begin{cases}
G_k \leq h_0 : & \text{restart the test at } k_o\leftarrow k+1\\
h_0< G_k < h_1 : & \text{monitor}\\
G_k\geq h_1 : & \text{declare detection}.
\end{cases}
\end{equation}
\end{subequations}
The thresholds $h_0$ and $h_1$ are selected to meet the desired false-alarm rate and detection-delay tradeoff. For the first time, the SLRT is started with $k_o=1$. With $h_0=0$, the SLRT becomes equivalent to the Page's test~\cite{Abraham2019} for detecting the signals with unknown start times.

\subsection{Simulation Setup}
To assess the statistical adequacy of the proposed background models and their impact on target detection, several time-varying underwater channel scenarios are simulated using BELLHOP~\cite{porter2011bellhop,Li2025}. Time variability is induced by (i) a rough sea surface synthesized from a wind-wave spectrum~\cite{hasselmann1973measurements} and (ii) random transmitter and receiver drift about their nominal positions, with independent horizontal-plane motion. Table~\ref{tab:bellhop_confg} summarizes the acoustic, environmental, and geometric parameters used to generate the received signal.

\begin{table}[t]
\centering
\caption{System and environment configuration used to generate the received signal.}
\label{tab:bellhop_confg}
\footnotesize
\renewcommand{\arraystretch}{1.05}
\resizebox{\columnwidth}{!}{%
\begin{tabular}{@{} l l @{}}
\toprule
\textbf{Parameter} & \textbf{Value} \\
\midrule
Signal type & Linear frequency-modulated (LFM) chirp \\
Number of pings & 100 \\
Pulse duration & 25\,ms \\
Pulse repetition interval (PRI) & 0.12\,s \\
Carrier frequency & 3\,kHz \\
Sampling frequency & 15\,kHz \\
Bandwidth (BW) & 4\,kHz \\
Ocean depth & 50\,m \\
Field generation mode & Time varying \\
Sound speed profile & Iso-velocity (1500\,m/s) \\
Bottom type & Acoustic half-space \\
Bottom density & 1.7\,g/cm$^{3}$ \\
Bottom attenuation & 0.5\,dB/$\lambda$ \\
Bottom roughness & 0.05 \\
Sound speed in bottom & 1650\,m/s \\
Distance between sensor nodes & 2\,km \\
Sensor node depth & 2\,m (from ocean surface) \\
Sea-surface spectrum & JONSWAP \\
Significant wave height $H_s$ & 1\,m \\
Peak period $T_p$ & 4\,s \\
Peak enhancement factor $\gamma$ & 3.3 \\
Node drift per axis $\sigma_{\text{pos}}$ & 0.20\,m \\
\bottomrule
\end{tabular}}
\end{table}

With this, three significant ocean scenarios are simulated
\begin{itemize}
    \item Scenario 1: a time-invariant baseline with a flat sea surface and fixed transmitter and receiver positions.
    \item Scenario 2: a surface-wave case with a moving surface and fixed transmitter and receiver positions.
    \item Scenario 3:  a fully time-varying case with a moving surface and drifting transmitter and receiver positions.
\end{itemize}

\noindent Figs.~\ref{fig:cir_bellhop_scenarios}(\subref{fig:cir_static})--(\subref{fig:cir_drift_reverb}) show the corresponding channel impulse responses (CIRs) for these oceanic scenarios.

To simulate a target return under these ocean scenarios, two target-motion conditions are considered:
\begin{itemize}
    \item A stationary target with fixed target delay.
    \item A moving target that crosses the bistatic baseline (transmitter-receiver line), producing a time-varying target delay.
\end{itemize}
The stationary target case is evaluated because both the multipath and the target exhibit similar Doppler values. Specifically, in the case of Scenario~1, the delay response of the target matches with the multipath background as shown in Fig.~\ref{fig:cir_bellhop_scenarios}(\subref{fig:cir_static}).

\begin{figure*}[t]
    \centering
    \begin{subfigure}[t]{0.24\textwidth}
        \centering
        \includegraphics[width=\linewidth]{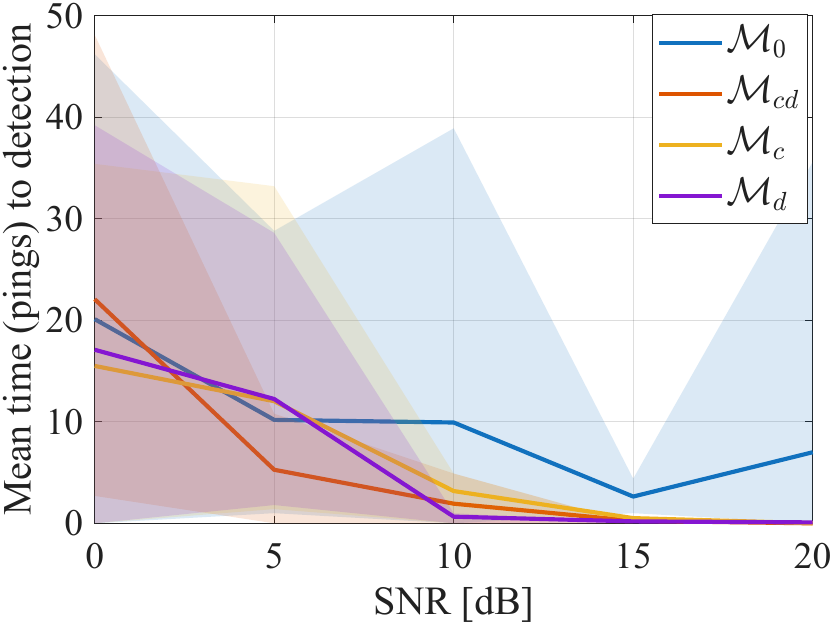}
        \caption{Stationary target, scenario~1.}
        \label{fig:mtd_s1_fixed}
    \end{subfigure}\hfill
    \begin{subfigure}[t]{0.24\textwidth}
        \centering
        \includegraphics[width=\linewidth]{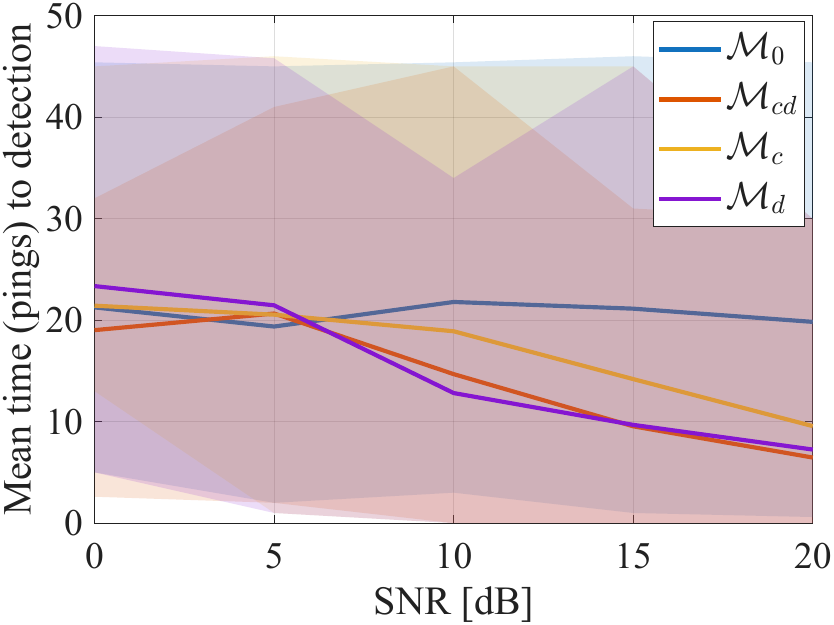}
        \caption{Moving target, scenario~1.}
        \label{fig:mtd_s1_moving}
    \end{subfigure}\hfill
    \begin{subfigure}[t]{0.24\textwidth}
        \centering
        \includegraphics[width=\linewidth]{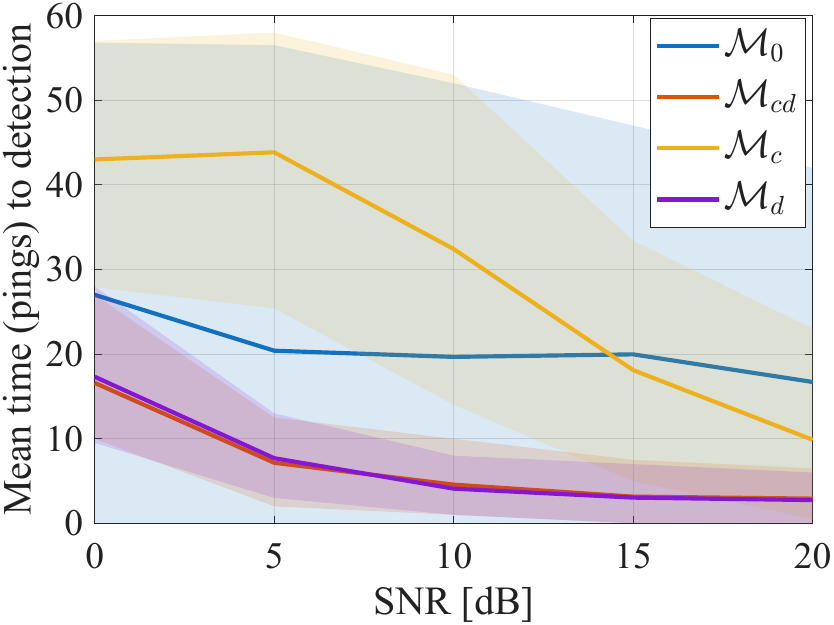}
        \caption{Stationary target, scenario~3.}
        \label{fig:mtd_s3_fixed}
    \end{subfigure}\hfill
    \begin{subfigure}[t]{0.24\textwidth}
        \centering
        \includegraphics[width=\linewidth]{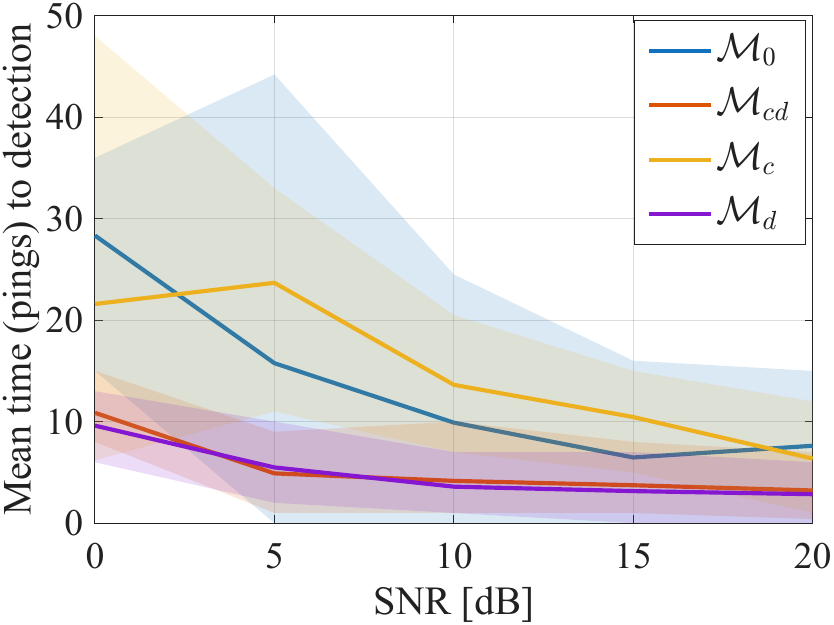}
        \caption{Moving target, scenario~3.}
        \label{fig:mtd_s3_moving}
    \end{subfigure}
    \caption{Mean time to detection (MTD) versus SNR at $\text{INR}=30$~dB. Solid lines show the mean detection delay over detected trials, and bounds indicate the 10--90 percentile range of detection delays.}
    \label{fig:mtd_vs_snr_all}
\end{figure*}
\begin{figure*}[t]
    \centering
    \begin{subfigure}[t]{0.24\textwidth}
        \centering
        \includegraphics[width=\linewidth]{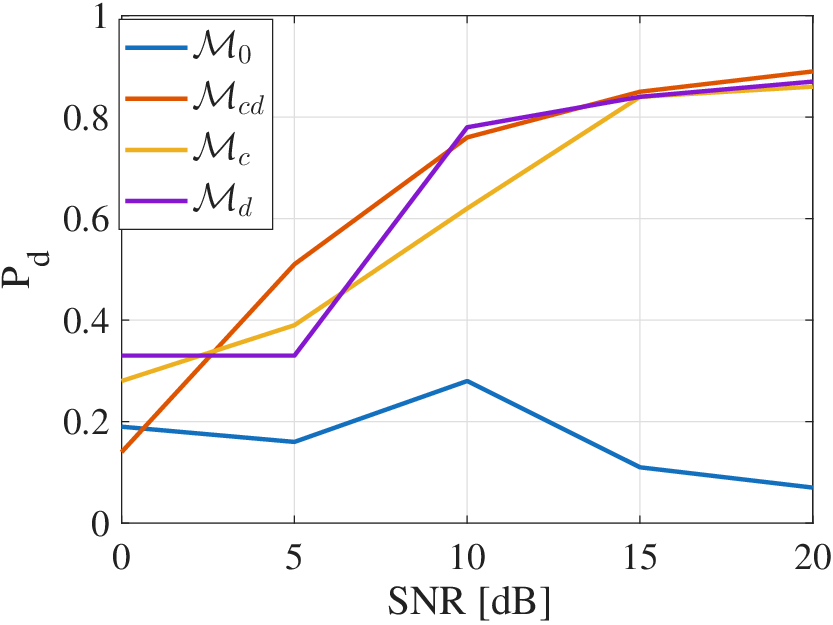}
        \caption{Stationary target, scenario~1.}
        \label{fig:pd_s1_fixed}
    \end{subfigure}\hfill
    \begin{subfigure}[t]{0.24\textwidth}
        \centering
        \includegraphics[width=\linewidth]{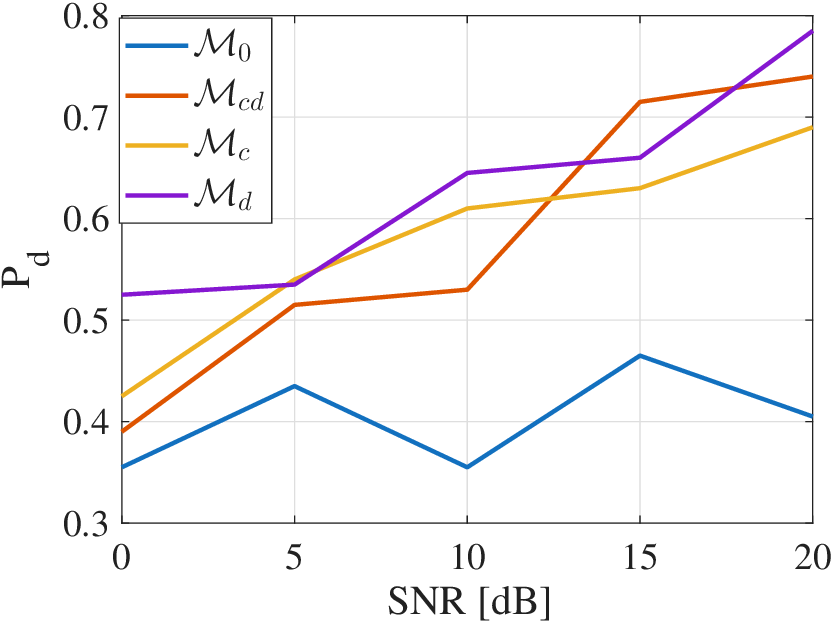}
        \caption{Moving target, scenario~1.}
        \label{fig:pd_s1_moving}
    \end{subfigure}\hfill
    \begin{subfigure}[t]{0.24\textwidth}
        \centering
        \includegraphics[width=\linewidth]{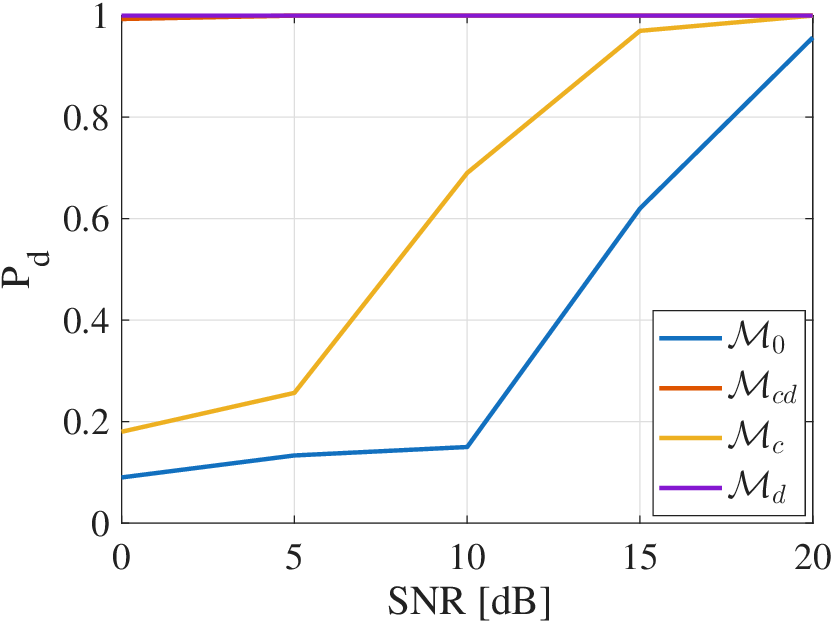}
        \caption{Stationary target, scenario~3.}
        \label{fig:pd_s3_fixed}
    \end{subfigure}\hfill
    \begin{subfigure}[t]{0.24\textwidth}
        \centering
        \includegraphics[width=\linewidth]{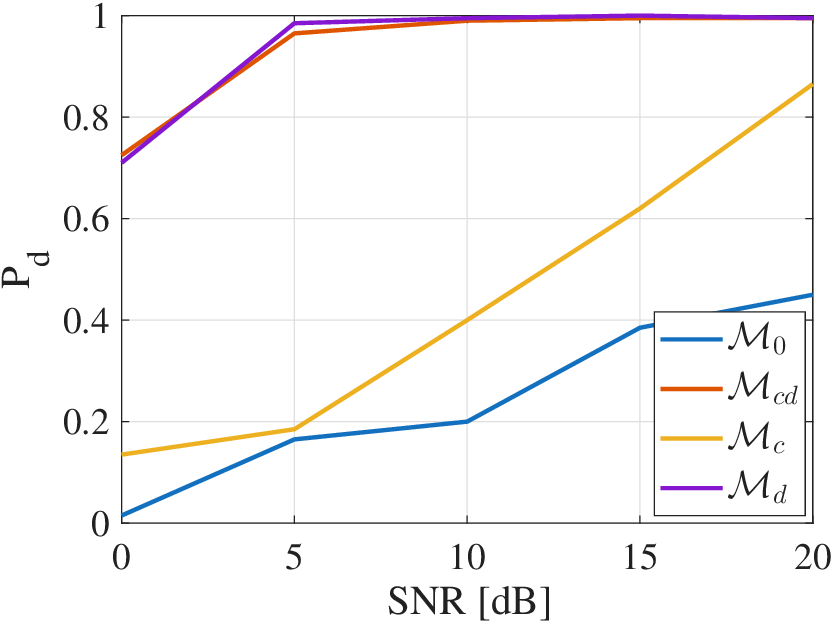}
        \caption{Moving target, scenario~3.}
        \label{fig:pd_s3_moving}
    \end{subfigure}
    \caption{Probability of detection $P_d$ versus SNR at $\text{INR}=30$~dB. A trial is counted as a missed detection if no alarm is raised by the final ping in the test horizon.}
    \label{fig:pd_vs_snr_all}
\end{figure*}

\begin{figure*}[t]
    \centering
    \begin{subfigure}[t]{0.24\textwidth}
        \centering
        \includegraphics[width=\linewidth]{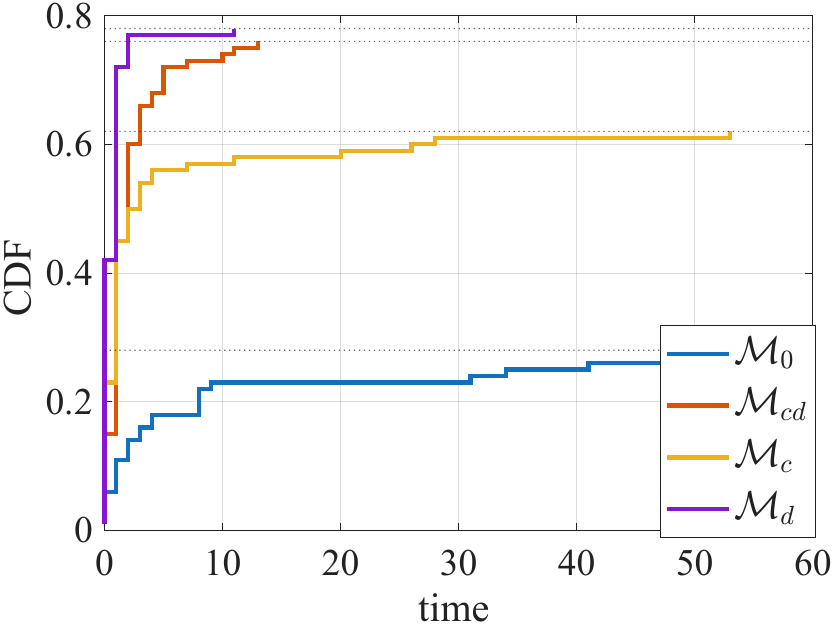}
        \caption{Stationary target, scenario~1.}
        \label{fig:cdf_s1_fixed}
    \end{subfigure}\hfill
    \begin{subfigure}[t]{0.24\textwidth}
        \centering
        \includegraphics[width=\linewidth]{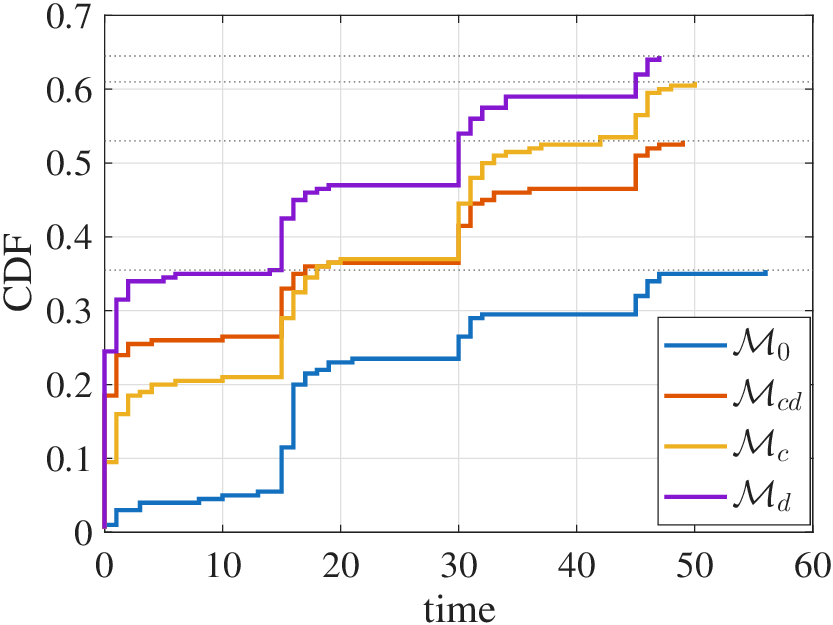}
       \caption{Moving target, scenario~1.}
        \label{fig:cdf_s1_moving}
    \end{subfigure}\hfill
    \begin{subfigure}[t]{0.24\textwidth}
        \centering
        \includegraphics[width=\linewidth]{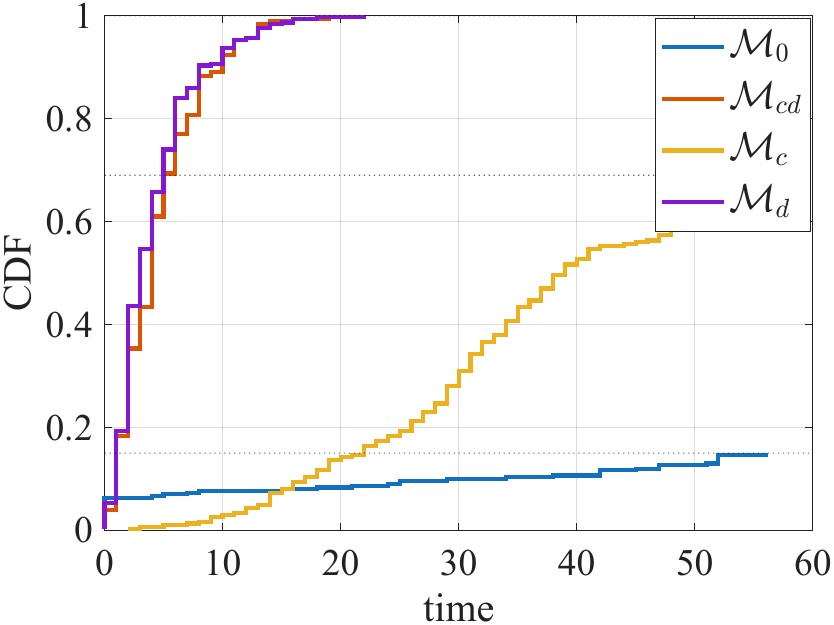}
        \caption{Stationary target, scenario~3.}
        \label{fig:cdf_s3_fixed}
    \end{subfigure}\hfill
    \begin{subfigure}[t]{0.24\textwidth}
        \centering
        \includegraphics[width=\linewidth]{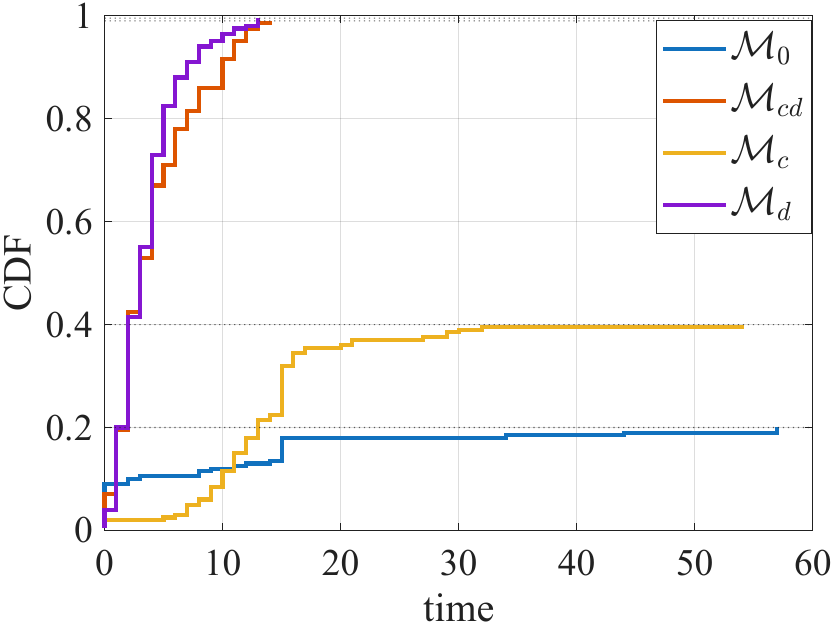}
        \caption{Moving target, scenario~3.}
        \label{fig:cdf_s3_moving}
    \end{subfigure}
    \caption{Cumulative distribution function (CDF) at an $\text{SNR} =10$~dB. The plateau level equals $P_d$, and the rise rate reflects detection latency.}
    \label{fig:cdf_delay_all}
\end{figure*}

\section{Results and Discussion}\label{sec:results}
With the simulated datasets, the statistical adequacy and target detection performance are evaluated across interference-to-noise ratio (INR) and signal-to-noise ratio (SNR), respectively, using $N_{\mathrm{MC}}$ Monte Carlo realizations. For each candidate background model in~\eqref{eq:models_clean}, a $p$-value test in~\eqref{eq: ptest} is evaluated. Significance is declared when $p\leq 0.05$.

Here, the INR and SNR are defined as
\begin{subequations}
\begin{align}
\text{INR}&=10\log_{10}\!\left(\frac{\|\vec{x}_{b}\|_2^2}{N\sigma_e^2}\right),\\
\text{SNR}&=10\log_{10}\!\left(\frac{\|\vec{x}_{o}\|_2^2}{N\sigma_e^2}\right).
\end{align}
\end{subequations}
The SLRT is applied to detect the target onset, which occurs after an initial target-free period. Specifically, the first $40$ pings (background-only) are used to learn the model hyperparameters $\Theta$, after which the sequential test is performed with the target introduced from ping $41$. The target detection performance is evaluated by the probability of detection $P_d$ as a function of SNR, mean time-to-detection (MTD) vs. SNR, computed only for target detections over $N_{MC}$ trials. Also, an empirical cumulative distribution function (CDF) is computed, which shows at which post-onset time delay index the target has been detected mostly. The SLRT threshold $h_1$ is calibrated empirically under $\mathcal{H}_0$, for all models and SNR values, to achieve a designed false-alarm probability $P_{\mathrm{fa}}=0.05$, and all reported $P_d$, MTD, and CDF results are estimated using $N_{\mathrm{MC}}=200$ trials.

Firstly, the significance of the proposed models is evaluated for INR from 0 to 30 [dB]. It is observed that in Scenario~1, $\mathcal{M}_c$ and $\mathcal{M}_{cd}$ are rejected for all tested INR. In contrast, in Scenarios~2 and 3, the extensions $\mathcal{M}_c$, $\mathcal{M}_d$, and $\mathcal{M}_{cd}$ yield $p$-values below $10^{-3}$ for essentially all INR values. This indicates substantial structure beyond $\mathcal{M}_0$ in the time-varying scenarios.

Even though the models in~\eqref{eq:models_clean} are low-dimensional approximations rather than exact models, the observed small $p$-values in Scenarios~2 and 3 suggest that the proposed covariance components capture structure beyond $\mathcal{M}_0$ in the simulated data.

The detection performance for the different background models is shown in Fig.~\ref{fig:mtd_vs_snr_all}, Fig.~\ref{fig:pd_vs_snr_all}, and Fig.~\ref{fig:cdf_delay_all}. The ambient noise-only baseline $\mathcal{M}_0$ yields near constant MTD (with a small decrease) and lower $P_d$ across the SNR sweep. It suggests that a homoscedastic measurement covariance is insufficient even under near-static conditions. With $\mathrm{INR}=30$~dB fixed, performance gains with increasing SNR primarily reflect improved separability between the target return and the learned background structure.  

The benefit of explicitly modeling the multipath background for the target detection is pronounced in all scenarios. Models that include the common-mode component, i.e., $\mathcal{M}_d$ and $\mathcal{M}_{cd}$, achieve consistently shorter MTD and higher $P_d$ for both stationary and moving targets. The CDFs corroborate this by showing earlier rises (faster detection) and higher plateaus (fewer misses) for the $\mathcal{M}_d$ and $\mathcal{M}_{cd}$ models.
Also, the analysis for Scenario~2 is not provided as there were marginal differences in performance for different models, and provided no extra insight, which is not captured by Scenarios~1 and~3.

Overall, the BELLHOP-generated results favor modeling the common-mode term parameterized by $\sigma_d$. Even when the model $\mathcal{M}_c$ provides only incremental improvement, accounting for common-mode fluctuations improves the target detection performance despite strong background interference. This is consistent with the observed CIR evolution in Scenarios~2 and~3, where the multipath structure changes slowly and coherently over time, making a drift-like background component more prominent.

\begin{comment}
\begin{figure}[t]
    \centering
    % --- Top row: mean delay to detection ---
    \begin{subfigure}[t]{0.5\columnwidth}
        \centering
        \includegraphics[width=\linewidth]{figs/Mdd_rnd_fixed.eps}
        \caption{Mean delay vs SNR.}
        \label{fig:mdd_s2_fixed}
    \end{subfigure}\hfill
    \begin{subfigure}[t]{0.5\columnwidth}
        \centering
        \includegraphics[width=\linewidth]{figs/Mdd_rnd_movingblock.eps}
        \caption{Mean delay vs SNR .}
        \label{fig:mdd_s2_moving}
    \end{subfigure}
    % --- Bottom row: Pd vs SNR ---
    \begin{subfigure}[t]{0.5\columnwidth}
        \centering
        \includegraphics[width=\linewidth]{figs/Pd_SNR_rnd_fixed.eps}
        \caption{$P_d$ vs SNR.}
        \label{fig:pd_s2_fixed}
    \end{subfigure}\hfill
    \begin{subfigure}[t]{0.5\columnwidth}
        \centering
        \includegraphics[width=\linewidth]{figs/Pd_SNR_rnd_movingblock.eps}
        \caption{$P_d$ vs SNR.}
        \label{fig:pd_s2_moving}
    \end{subfigure}
\caption{SLRT performance for scenario~3 with mean delay to detection (top) and terminal $P_d$ (bottom) versus SNR for a stationary target in (a), (c), and a moving target in (b), (d).}
    \label{fig:seq_page_performance_s3}
\end{figure}
\end{comment}

\section{Conclusion}
To summarize, the presented results indicate that the proposed wide-band Doppler linearization-based model can reliably describe Doppler effects due to platform and surface wave motion. This enables efficient background learning in the raw acoustic measurement domain to bolster the detection of weak target(s) embedded in the multipath background. Future research will focus on integrating the model into a track-before-detect framework for joint background learning, target detection, and tracking.

\appendices
\section{Wideband linearization of time-varying multipath signal}\label{app:linearization}
Define $F(r,t)\triangleq s(e^r t)$ with $r=\mathrm{ln}\beta$. Then,
\begin{equation*}
    \frac{\partial F(r,t)}{\partial r}= e^r t\dot{s}(e^r t) =t\frac{d F(r,t)}{dt}
\end{equation*}
With $F(0,t)=s(t)$, the solution to the PDE is
\begin{equation}\label{app:eq1}
    F(r,t)=e^{r(t\frac{d}{dt})}s(t), \quad s(\beta t)=e^{\mathrm{ln\beta}(t \frac{d}{dt})}s(t).
\end{equation}
Using the series expansion,
\begin{equation*}
    e^a=1+a +\frac{a^2}{2!}+\frac{a^3}{3!}+\cdots
\end{equation*}
and taking the first order approximation,~\eqref{app:eq1} can be written as
\begin{equation}
    s(\beta t) \approx s(t) +\mathrm{ln}\beta t \dot{s}(t).
\end{equation}

\IEEEtriggeratref{2}
\bibliographystyle{IEEEtran}
\bibliography{IEEEabrv,FUSION2026}

\end{document}